# Complexity measures and occurrence of the "breaking point" in the neutron and gamma-rays time series measured with organic scintillators

D. Mihailovic[a,*], S. Avdic[b] and A. Mihailovic[c]

[a]Department of Physics, Faculty of Sciences, University of Novi Sad, Dositeja Obradovića Sq., 3, 21000 Novi Sad Serbia

email: guto@df.uns.ac.rs

[b] Department of Physics, Faculty of Science, University of Tuzla, Univerzitetska 4, 75000 Tuzla, Bosnia and Herzegovina

email: senada.avdic@untz.ba

[c] Faculty of Technical Sciences, University of Novi Sad, Dositeja Obradovića Sq., 1, 21000 Novi Sad, Serbia

email:mihailovic.dt2.2020@uns.ac.rs

[*]Corresponding author. D.T. Mihailovic Department of Physics, Faculty of Sciences, Dositeja Obradovića Sq., 3, 21000, Novi Sad, Serbia

email: guto@df.uns.ac.rs

**Abstract**

This paper deals with the first analysis of the neutron and gamma time series measured with organic scintillators from plutonium samples by using information measures. Fast neutron detection with organic scintillators has been widely used for various nuclear safeguards applications and homeland security. One of the significant attributes of special nuclear materials (SNM) is the high multiplicity events in a short period of time. The time distributions of neutron and gamma-rays events for the plutonium metal plates designed as fuel plates for the Idaho National Laboratory (INL) Zero Power were measured with the Fast Neutron Multiplicity Counter (FNMC) consisting of 8 EJ-309 liquid scintillators and 8 stilbene detectors. Since the neutron correlated counts within the coincidence window of 40 ns are related to $^{240}$Pu effective mass of plutonium metal plates, it is of interest to investigate the randomness of the measured neutron and gamma-rays events. To access such information, we resort to complexity measures in the hope of being able to connect complexity values with the reliability of detection. That was done through (i) application of Kolmogorov complexity (KC) and its derivatives [Kolmogorov spectrum and its highest value (KCM) and running complexity (RKC)] and (ii) establishing the "breaking point" after which there exists a sharp drop in the running Kolmogorov complexity of the neutron and gamma-rays time series. It was found that the complexity of all the time series detected from the sample with 5, 9, 11, 13, and 15 plutonium plates had the high almost identical values of KC while the sample with 3 plates had by one-third smaller KC values than all the others. These calculations were supplemented by the Lypaunov exponents for a time series and the NIST tests. The low KC values can be addressed to the different sources of uncertainties in measuring procedure with the sample consisting of three plates. The complexity measures applied in this study are capable of revealing aspects of information that would otherwise remain hidden to the one-off complexity estimate.

## 1. Introduction
### 1.1 Detection of special nuclear materials

Detection of special nuclear materials (SNM) such as plutonium, uranium-233, or uranium-235 (undeclared material and the missing of declared material) is of high importance in the field of nuclear safeguards and homeland security (Robinson et al., 2011). Passive neutron coincidence counting is the frequent technique for the non-destructive assay (NDA) of SNM. Detection systems based on organic scintillators have been applied successfully in recent nuclear safeguards applications for measurements on the nanosecond time-scales (Knoll, 2010). Taking into account that the neutron lifetime within organic scintillators is several orders of magnitude shorter compared to the neutron lifetime in thermal neutron counters it is possible to reduce the coincidence window significantly and achieve lower statistical measurement uncertainty compared to thermal neutron systems (Chichester, 2015). SNM emits multiple neutrons and gamma-rays simultaneously, through spontaneous fission (SF) or neutron-induced fission (IF) reactions. Since signatures for fissile materials are the high multiplicity events in thenanosecondtime-scale, detection systems based on organic scintillators are suitable for the detection of spontaneous and neutron-induced fission events. Neutrons from the spontaneous

fission,(alpha, n) reactions and background, induce fissions, and generate some very short-lived chains in most cases (Pazsit et al., 2015).

For this study we used the measured neutron and gamma-rays time series for 3, 5, 7, 9, 11, 13, and 15 PAHN Pu metal plates. The measurements were performed by using the Fast Neutron Multiplicity Counter (FNMC) consisting of 8 EJ-309 liquid organic scintillators and 8 stilbene detectors at Idaho National Laboratory (INL) (Di Fulvio, et al., 2017). Both types of organic scintillators are sensitive to fast neutrons and gamma rays. The pulse-shape discrimination (PSD) between fast neutrons and gamma rays was carried out using a standard charge integration method (Polack et al., 2015). The neutron time interval distributions depend on the source activities providing a possibility to evaluate $^{240}$Pu effective mass of the samples with high accuracy (Di Fulvio et al., 2017). Since the neutron time distribution of events detected with the FNMC in a coincidence acquisition mode can be useful for extracting some fissile material properties such as the fissile material mass, it is of interest to investigate the randomness of neutron and gamma-rays events in the process of their detection.

1.2 Studying the complexity of neutron and gamma-rays time series

Understanding the dynamics of nuclear fission remains a challenge in nuclear physics. Nuclear fission is a stochastic process and does not always take place in the same way, which means that two identical fission fragments are not obtained with a high probability or that the same number of neutrons and gamma-rays are always emitted. Spontaneous fission within the coincidence window of 40 ns is related to the fissile material massand one of the ways to better understand the dynamics of SF process is to measure the complexity of their time series. The assumption is that complexity represents an important variable that comes from the intuitive understanding of the unpredictability of chaos in radioactive decay (SF can be treated as a form of radioactive decay emitting multiple neutrons and gamma-rays in a short time interval) and the high complexity and low entropy of the measured time series. To access such information, we resort to complexity measures in the hope of being able to connect complexity values with various external causes and reliability of detection.

The structure of the paper is as follows. Section 2 presents the clear mathematical consideration and description of Kolmogorov complexity and its derivatives (subsection 2.1) that includes Kolmogorov complexity (2.1.1), Kolmogorov spectrum and its highest value (2.1.2) and running complexity (RKC), and "breaking point" of the running complexity time series (2.1.3), in which we defined this term by means of numerical experiments with a logistic equation (LE). Subsections 2.4 and 2.5 are devoted to description of statistical randomness and NIST tests and calculating the Lyapunov exponent from a time series.

In Section 3 we describe the experimental set up that includes (i) technical details and (ii) acquisition and processing of Pu time series (subsections 3.1 and 3.2, respectively). The results are discussed and conclusions are given in Section 4.

## 2. Methods

Experiment is one of the crucial methods in physics. Paraphrasing Duhem (1954), it could be said that an experiment can always remove a theory from power, but never prove it definitively. In this paper, we deal with occurrence of the "breaking point" in the neutron and gamma-rays time series detected by organic scintillators. On that occasion, we use the term randomness, which have different interpretations. Therefore, we will consider its possible interpretations and explain why we decided on the one that will be described below in details.

If we ask ourselves a question: what is randomness and where it comes from, then we must be aware that "this is one scary place to venture in" (Perez, 2017). We take for approved the randomness in our reality (consciously or unconsciously; with or without intent). To avoid this we use the probability theory to compensate for the randomness. In a really enlightening paper "Introduction to foundations of probability and randomness" about the randomness, Andrei Khrennikov explored the different interpretations of randomness. He also elaborated in more detail Kolmogorov's proposal on this subject (Khrennikov, 2014). He summarizes the three different interpretations of randomness: (1) Randomness as unpredictability; (2) randomness as typicality and (3) randomness as complexity. He observed that: "As we have seen, none of the three basic mathematical approaches to the notion of randomness (based on unpredictability, typicality, and algorithmic complexity) led to the consistent and commonly accepted theory of randomness." Among these interpretations, only Kolmogorov's interpretation is of an objective nature. To our surprise, this view point is not actually the predominant viewpoint in our understanding of probability. Moreover, Kolmogorov complexity is not even computable. Therefore, randomness is either a subjective measure or a non-computable objective measure. According to his discussion it might be that randomness is not mathematical, but physical notion. Namely, it is the physical procedure where the true randomness is hidden. It is plausibly possible that strictly mathematical approach to randomness and formalization of its definition cannot be reached, since probably it is insufficient to use only the mathematical methods for the theoretical construction of concept of the randomness itself. Let it note that in Kolmogorov's interpretation there is "*there is no room high enough*" for *level* of randomness.

Many scientific analyzes use the term "algorithmic" randomness, which is directly related with the definition of complexity developed by Kolmogorov (Kolmogorov, 1965). It is quantified by the algorithmic complexity, which is a measure of how long an algorithm would take to complete given an input of size *n*. If an algorithm has to scale, it should compute the result within a finite and practical time bound even for large values of *n*. For this reason, complexity is calculated asymptotically as *n* approaches infinity as a measure of randomness. Note that a mistreatment of language allows us to attribute greater randomness to a sequence with a higher complexity value. However, we maintain the term level in sense of the above approach. This is a typical situation in science (in particular, in physics and engineering sciences). Namely, in choosing the model or approach, scientists often apply a *heuristic* technique that could be defined as any approach to problem solving that makes use of a practical method not guaranteed to be optimal or perfect but sufficient either for the immediate goals or until a better approach is reached.

**2.1. Kolmogorov complexity and its derivatives**

**2.1.1. Kolmogorov complexity**

The Kolmogorov complexity (KC) can be defined as the complexity of an object, such as a piece of text, to be a measure of the computability resources needed to specify the object. This complexity in existing terms is the minimum piece of code/program that one can write to generate a particular string. This measure is non-computable. Therefore, for a given time series, $x_i$, $i = 1,2,3,4,...,N$ it is calculated approximately using the Lempel-Ziv algorithm (LZA, Lempel and Ziv, 1976), or some of its variants. This algorithm includes the following steps: (1) Encode the time series by constructing a sequence $S$ of the characters 0 and 1 written as $s(i), i = 1,2,...,N$ according to the rule $s(i) = 0, x_i < x_m$ or 1 if $x_i > x_m$, where $x_m$ is the mean value of the time series samples, selected as the threshold (Mihailović et al., 2015). Other encoding schemes are also used, depending on the application (Radhakrishnan et al., 2000). (2) Calculate the complexity counter $C(N)$. The $C(N)$ is defined as the minimum number of distinct patterns contained in a given character sequence. The complexity counter $C(N)$ is a function of the length of the sequence $N$. The value of $C(N)$ is approaching an ultimate value $b(N)$ as $N$ is approaching infinity, i.e., $C(N) = O(b(N))$ and $b(N) = N/log_2 N$. (3) Calculate the normalized information measure $C_k(N)$ which is defined as $C_k(N) = C(N)/b(L) = C(N) log_2 N /N$. The parameter $C_k(L)$ represents the information quantity contained in a time series, and it varies from 0 to 1, although the values of KC can exceed value 1.

**2.1.2 Kolmogorov complexity spectrum and its highest value**

The Kolmogorov complexity has two weaknesses: (i) It cannot distinguish between time series with different amplitude variations and that with similar random components; and (ii) in the conversion of a time series into a binary string, its complexity is unseen in the rules of the applied procedure. Therefore, in defining a threshold for a criterion for coding, some information about the composition of time series could be lost. In our analysis, two measures were used: (i) Kolmogorov complexity spectrum (KC spectrum) and (ii) the highest value of the KC spectrum, introduced by Mihailović et al. (2015) who described the procedure for calculating the KC spectrum. Figure 1 schematically shows how to calculate the KC spectrum $C(c_1, c_2, c_3, ..., c_N)$ for plutonium (neutron and gamma-rays) time series $X(x_1, x_2, x_3, ..., x_N)$. This spectrum allows investigating the range of amplitudes in a time series that represents a complex system with highly enhanced stochastic components. The binarization of time series reflects an "average" since it depends on what value is used to binarize the series. This problem is also present in statistical randomness tests. However, the KC spectrum takes this problem into account, since it calculates the complexity taking each element of a time series as a threshold. Ultimately, this relates probability distribution with compressibility.

It may be noted that for a large number of samples of a time series, the computation of the KC spectrum can be computationally consumable. Therefore, it is reasonable to divide the domain, including all values between the minimum ($c_{min}$) and maximum ($c_{max}$), in the time series into subintervals, which then are used as thresholds. The highest value $K_m^C$ as in this series, i.e., $K_m^C = max\{c_i\}$, is the highest value of the Kolmogorov complexity spectrum (KCM), as seen in Figure 1.

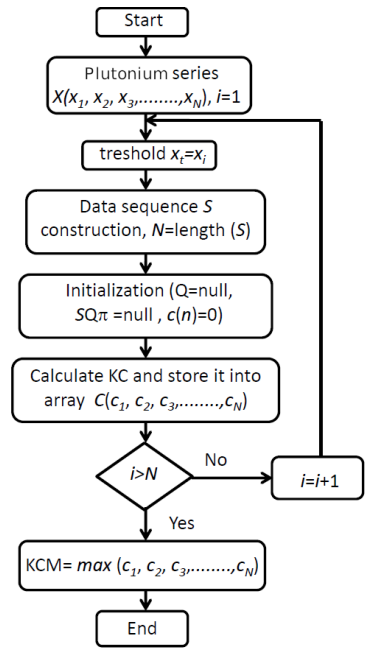

**Figure 1** Flow chart for calculation of the Kolmogorov complexity spectrum and its highest value (KCM) (Mihailovic et al., 2019).

**2.1.3 Running Kolmogorov complexity and "breaking point"**

In nuclear statistics, a running mean is often calculated to analyze data by creating a time series of averages of different subsets of the full data set that is a type of finite impulse response filter. This procedure was used to calculate the Running Kolmogorov complexity (RKC). For a given time series of plutonium samples, we extracted a fixed window as in running mean procedure, and then the procedure for calculating the KC (Fig. 1) is applied. After that, the window is moved forward for a step, and the KC algorithm is applied, until the end of Pu, the time series is reached. Note that the running temporal means offers a consistent and easily interpretable summary of the time series not only for KC but also for other measures.

On epistemological grounds, it is quite disturbing that there exists randomness to nature. We mentioned the word randomness. In everyday language but also in some scientific branches, people tend to use the words "random" and "chaotic" synonymously. Here, we make a clear distinction between their meanings. The *chaotic* sequences are in fact generated deterministically from the dynamical system $x_{n+1} = f(x_n)$ where $f$ is a smooth function of $\Re^m$. A bounded sequence of values $\{x_i\}_{i=1}^{\infty}$ coming from the previous equation is chaotic if (1) $\{x_i\}$ is not asymptotically periodic; (2) No Lyapunov exponent ($\lambda$) vanishes and (3) the largest Lyapunov exponent is strictly positive. *Random* (also called *stochastic*) as a result will give different sequences, even if the initial state is the same.

Here, we will show computationally that in the condition of high chaos and randomness, the KC complexity can unexpectedly go down or even more to enter an increasing mode. For this purpose, we will use the logistic equation (LE) that deterministically generates chaotic sequences (logistic parameter $r$ greater than 3.57). One of the indicators of chaotic behavior is the Lyapunov exponent whose positive value indicates the presence of chaos. We generated six time series, each of length 3000, covering an interval of $r$ from 3.948 to 3.953 with an increment of 0.001. Then we calculated the running mean of Lyapunov exponent for a fixed window (of sizes 300, 400, 500, and 2800) for each $r$ (Figure 2a). From this figure is seen that all values of $\lambda$ are positive (most of them indicate a high chaotic regime).

The same experiments were performed for KC complexity. For the shortest window, a size of 200 steps was selected according to numerical experiments, which indicate that it is the minimum length at which calculating the KC complexity, by the LZA, is applicable (Mihailović et al., 2014). The results of the experiments are given in Fig. 2b. For some values of the logistic parameter the randomness of the sequence, quantified by KC complexity, at some point starts to go down. We will define that point as a "breaking point" of the running KC complexity (or shortly "breaking point") for randomness.

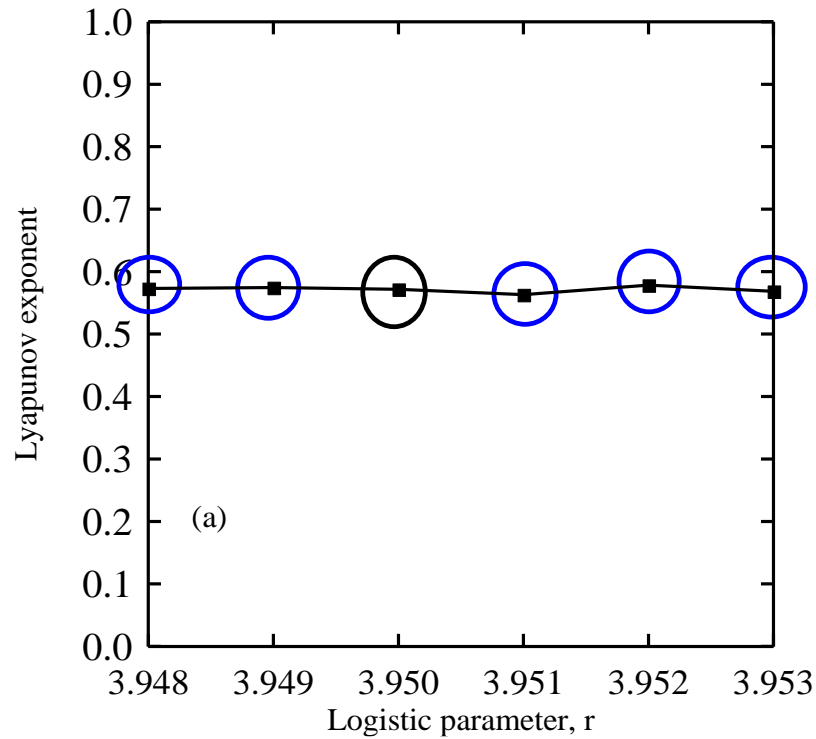

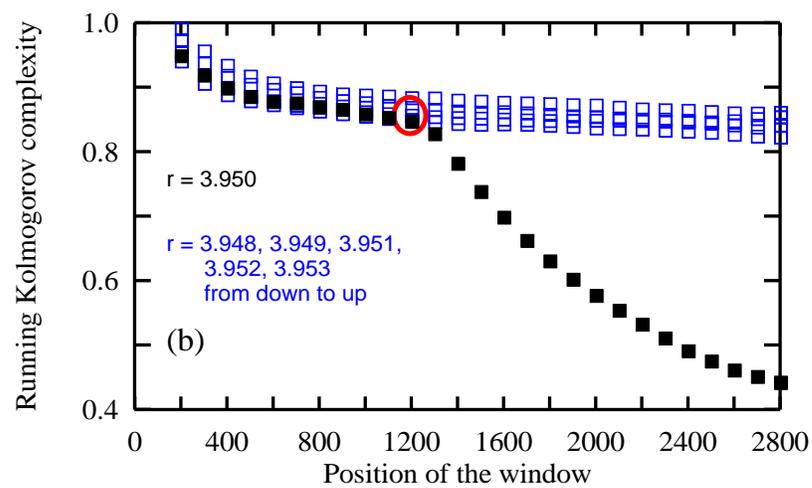

**Figure 2** The running mean of Lyapunov exponent of the LE (a) and KC complexity (b) time series for a fixed window (of sizes 200, 300, 400, … , 2800). Calculations were performed for six-time series, each of length 3000, covering an interval of *r* from 3.948 to 3.953 with an increment of 0.001. Black circle surrounds the value of *r* this parameter used in the experiment. Red circle indicates position of the "breaking point".

### 2.2. Statistical randomness

A numeric sequence is said to be *statistically random* when it contains no recognizable patterns or regularities; sequences such as the results of an ideal die roll, or the digits of Pi or some of other the best known transcendental numbers exhibit "statistical randomness. It is one contrast to the algorithmic randomness which *per definitionem* does not use probabilities. A sequence exhibiting a pattern is not thereby proved not statistically random. Tests for statistical are often used in science, in particular in analysis of outputs in quantum mechanics and other fields of experimental physics. For example, Kovalsky et al (2018) analyzed both complexities of sequences of random numbers generated in Bell's experiments. To analyze gamma and neutron time series we used the battery of tests provided by the National Institute of Standards and Technology (NIST). However, The NIST tests (Shen et al., 2016) are quite expensive in computational time requiring very long series. Therefore we used a version that is less time consumable (Sys et al., 2015). The complete battery includes 15 tests but we used the simplest 6 following Kovalsky et al (2018). They were: Frequency (Monobit) Test, Frequency Test within a Block, Runs Test, Tests for the Longest-Run-of-Ones in a Block, Binary Matrix Rank Test and Discrete Fourier Transform (Spectral) Test. A run is said to have positive ("yes") randomness only if passes the 6 tests.

## 2.3. Determining Lyapunov exponents from a time series

Since a random process is non-deterministic, numerical computation of a "Lyapunov exponent" is not well defined but. Wolf's algorithm (applied to a sequence) looks at the closest neighbor $y_0=x_K$ of a point and uses it as the initial value of another sequence. The average value of the rate of separation $\ln(\|x_{K+1} - x_{J+1}\|)/\|x_K - x_J\|)\|x_K - x_J\|)$ is used as a measure of the Lyapunov exponent. Calculations were performed following Kodba et. al. (2005) with software which is available on http://www.matjazperc.com/time/ which is based on Wolf's algorithm that is broadly used although it is also designed on a heuristic way.

## 3. Experimental setup

### 3.1 Technical details

#### 3.1.1 Experimental configuration

Experimental validation of the FNMC measurement system consisting of sixteen organic scintillation detectors was performed at Idaho National Laboratory (INL) (Di Fulvio et al., 2017).The 3D model of the FNCM is given in Figure 3. A passive assay in a coincidence acquisition mode of plutonium metal plates (3, 5, 7, 9, 11, 13, and 15 PAHN plates) was carried out in the mass range between 75.14 g and 375.69 g. The experiments were done with lead shielding to reduce the gamma-ray background for a symmetric position of the plutonium samples with respect to the detectors.

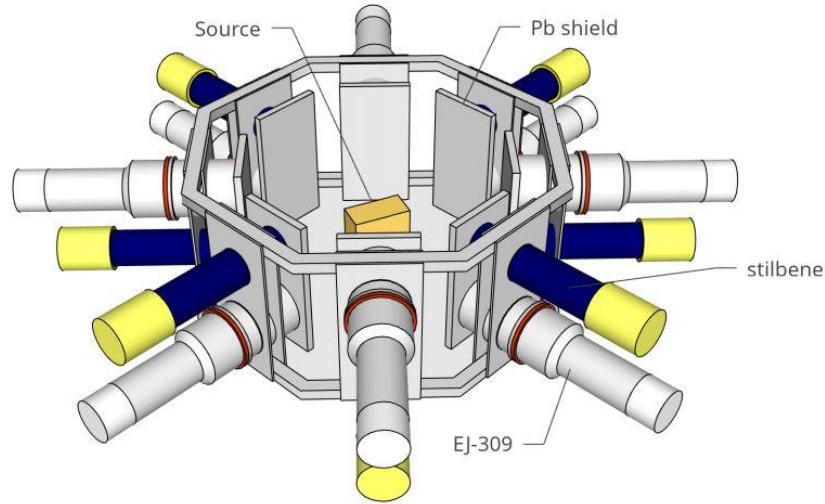

**Figure 3** Three dimensional model of the FNMC measurement system.

The FNMC includes 8 EJ-309 liquid scintillation detectors with 7.62-cm diameter by 7.62-cm length (by Eljen, Sweetwater, Tex.) and 8 stilbene detectors with 5.08-cm diameter by 5.08-cm length (by Lawrence Livermore National Laboratory and Inrad Optics, Northvale, N.J.). Both types of scintillators have similar composition and good capability of PSD which is of high importance for the safeguards applications with the high gamma-ray background. Analgorithm for charge-integration, pulse-shape discrimination. and estimation ofneutron/photon misclassification in organic scintillators was applied in the measurements (Polack et al., 2015). Since stilbene detectors have better PSD properties, a wider dynamic range of 336–6500 keV neutron energy deposited for stilbene detectors was used in the experiment compared to a range of 520–5530 keV neutron energy deposited for EJ-309 detectors. The digital acquisition was carried out by using a 500-MS/s, 14-bit, 16-channel digitizer (V1730 by CAEN Technologies, Viareggio, Italy (S.P.A., 2016).

### 3.1.2 Properties of plutonium metal plates

A set of fuel plates designed and built for the INL Zero Power Physics Reactor (ZPPR) was used as the samples in the measurements with the FNMC. At the date of the experiment PAHN plates, with a total mass of approximately 100 g for each plate, contained 23.92±1 g of $^{240}$Pu and 79.69±2 g of $^{239}$Pu. Isotope mass uncertainties are given taking into account the known uncertainty in the half-life of each isotope. The exact isotopic composition of the PAHN plutonium plates with the external dimensions of 7.62 cm by 5.08 cm and thickness of 0.3175 cm with a thin 304 L stainless steel cladding for encapsulation of the PAHN plutonium metal core is given in Table 1. The time interval distributions for the PAHN plutonium plates obtained with the FNMC in a 3 min assay time for all the samples were used for estimation of $^{240}$Pu$_{eff}$ mass of the plutonium samples (Di Fulvio et al., 2017) while the present study is focused on the time series analysis by the information measures.

*Table 1 Isotopic composition of the PAHN Pu plates with isotope mass uncertainty, (aged to the date of the experiment, Di Fulvio et al., 2017).*

| Nuclide | PAHN mass (g) |
|---|---|
| $^{238}$Pu | 0.00020(7) |
| $^{239}$Pu | 79.69(2) |
| $^{240}$Pu | 23.92(1) |
| $^{241}$Pu | 0.65(1) |
| $^{242}$Pu | 0.67(1) |
| $^{241}$Am | 1.87(1) |
| Al | 1.25 |
| Total plutonium | 106.79 |

**3.2 Acquisition and processing of Pu (neutron and gamma-rays) time series**

The time interval distributions of neutrons and gamma rays from all the PAHN Pu plates were acquired in the same time interval of 3 min. In the light of the foregoing, the Pu (neutron and gamma-rays) time series are of different lengths for the PAHN Pu plates having various amounts of plutonium. Taking into consideration (i) that neutrons emitted in spontaneous fission are correlated in a short time interval and (ii) that the correlated neutron counts are related to $^{240}$Pu effective mass of plutonium PAHN Pu plates (Di Fulvio et al., 2017), the Pu time series were analyzed within a short coincidence time window of 40 ns. In Table 2 are given the lengths for each time series measured by 8 stilbene scintillators and 16 (8 EJ-309 and 8 stilbene) scintillators.

*Table 2 Number of samples in the Pu time series.*

| Number of Pu plates | 8 detectors | | 16 detectors | |
|---|---|---|---|---|
| | Neutrons | Gamma rays | Neutrons | Gamma rays |
| | Number of samples | Number of samples | Number of samples | Number of samples |
| 3 | 8 028 | 72 696 | 63 309 | 423 018 |
| 5 | 9 810 | 84 532 | 77 359 | 362 963 |
| 7 | 15 620 | 131 528 | 125 343 | 472 881 |
| 9 | 22 101 | 182 820 | 178 435 | 546 115 |
| 11 | 61 316 | 485 974 | 486 122 | 1 315 164 |
| 13 | 79 765 | 620 895 | 630 348 | 1 525 240 |
| 15 | 99 031 | 767 432 | 791 463 | 1 791 550 |

It is obvious that due to the large difference in series lengths, the question can be asked to what extent the results for KC calculated using the LZA algorithm are representative of intercomparison. There is no general conclusion about the dependence of KC on the length of natural time series. In that regard, we relied on our experience in working on (i) logistic equation (Mimić et al., 2016), (ii) time series of velocity data of turbulent flow in laboratory channel (Mihailović et al., 2017), and river streamflow (Mihailović et al., 2019). The number of samples in those experiments varied between 3 000 and 9800. These experiments have shown that the running complexities of turbulent time series slightly increase with the time series length,and achieve saturation at some point. Thus, the difference between its highest and saturated value is less than 7%. Similar results were found in experiments with river streamflow time series. The saturation of running complexities in the time series generated by the logistic equation is reached after a remarkably smaller length. These results indicate that the dynamic of the measured time series is more complex than the dynamic of deterministically created a chaotic time series.

The RKC for the PAHN Pu three plates (PAHN_3) time series of neutron and gamma values measured by 8 stilbene scintillators and 16 scintillators (8 EJ-309 and 8 stilbene detectors) were calculated using the windows with length of 200 time units. As in the running mean procedure, the procedure for calculating the KC (Fig. 1) was applied. After that, the window is moved forward for a step with a length of one-time unit and the KC algorithm is applied, until the end of the Pu time series is achieved. In this study, 500 thresholds were used to obtain all spectra.

Results reported here were computed in FORTRAN90 while graphics was done in MATLAB.

**4. Results and discussion**

In a nonlinear dynamic analysis of neutron and gamma-rays time series from the metal Pu samples measured with organic scintillators, we encounter two sources of randomness that come (i) from nature these time series and (ii) background of scintillators. All the measured time series have high randomness and also elements of chaotic behavior, thus their analysis, like astrophysical and cosmic rays time series is concerned with time-varying signals that contain a non-trivial element of randomness (Vaughan, 2013). The changes in that randomness can occur because of uncertainties and characteristic limits of organic scintillators.

**4. 1 Measuring evidence**

Plutonium as a SNM emits multiple neutrons and gamma rays simultaneously and therefore they are correlated in time. Common measures for evaluation of the correlation between the detected particles include cross-correlation in time domain. Cross-correlation functions are based on differences between the arrival times of two correlated detections (Clarke, et al., 2009).The neutron time interval distributions of the samples are produced by spontaneous fissions, induced fissions, and background events [neutrons from (alpha, n) reactions in plutonium metal plates were negligible]. The coincident detections can be classified either as real or accidental coincidence. Real coincidences occur when multiple particles from the same reaction are detected in the coincident window whereas accidental coincidences occur when two particles from the different reactions are detected in the coincident window. The total correlations are related to the total number of counts due to real and accidental correlations in the entire measurement window. Cosmic rays can create a neutron background that can also initiate fission chains. A significant contribution of cosmic ray events to the interpretation of the signal was determined in the case of passive neutron counting of high enriched uranium (HEU) samples of low multiplication (Nakae et al., 2011). The contribution of accidental coincidences due to the source and random background events is not negligible in the measurements with fast organic scintillators. Taking into accout that fission emits multiple neutrons simultaneouly, the most of (neutron, neutron) coincidences can be attributed to fission reaction with a low but measurable contribution of accidentals.

The number of neutrons emitted in spontaneous fission of $^{240}$Pu varies from zero to six or more whereas the number of neutrons emitted in induced fission (IF) of $^{239}$Pu can vary from zero to eight. Neutron multiplicity distributions for $^{240}$Pu and $^{239}$Pu are given in Figure 4. The average SF neutron and gamma-rays multiplicity in $^{240}$Pu is about 2.156 (Ensslin et al., 1998) and 8.2 (Oberstedt et al., 2016) respectively, while the average IF neutron multiplicity in $^{239}$Pu is about 3.16 induced by neutrons with energy of 2 MeV (Zucker and Holden, 1986). The spontaneous fission (SF) rate is related to the effective $^{240}$Pu mass in the PAHN Pu samples, whereas the induced fission rate is proportional to the content of the odd-numbered fissile isotopes. The common definition of the $^{240}$Pu$_{eff}$ mass is related to the mass of $^{240}$Pu that would give the same response in terms of neutron coincidences as that obtained by the actual $^{238}$Pu, $^{240}$Pu, and $^{242}$Pu content of the sample.

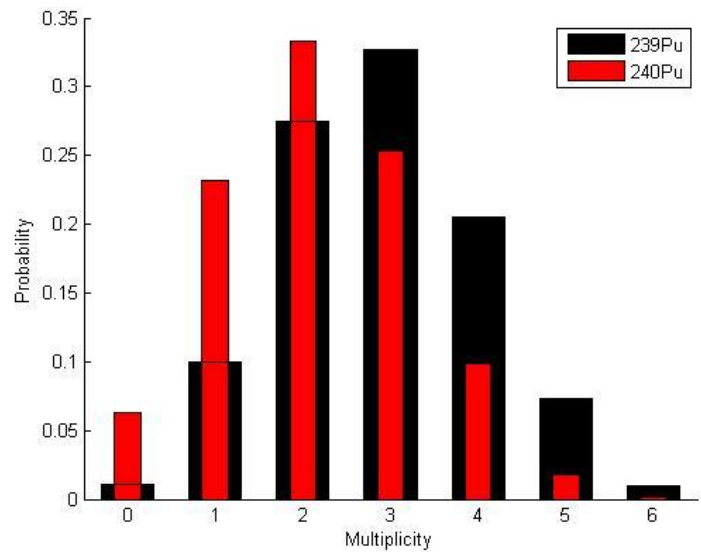

**Figure 4** The SF neutron multiplicity distribution for $^{240}$Pu and the 2-MeV-neutron-induced fission multiplicity distribution for $^{239}$Pu (Ensslin et al., 1998).

Figure 5shows the logarithm of the time interval between neutron events plotted against the running time on the linear x-axis for 3 and 15 PAHN Pu plates. It can be noticed in the range from 40 ns to 10 µs that the number of induced neutrons is much higher for the sample with more $^{239}$Pu (15 PAHN plates) compared to the sample with 3 PAHN plates. This is clear experimental evidence of the presence of SNM since the counts at this intermediate time scale can only occur due to neutrons inducing fission in the Pu samples.

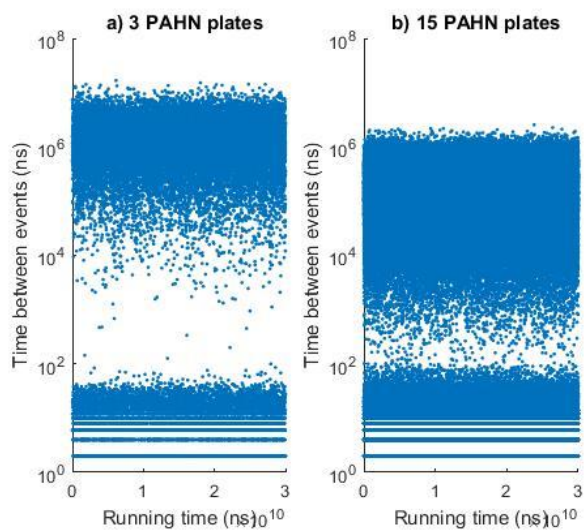

**Figure 5** Distributions of the time interval between detected neutrons as a function ofthe running time for PAHN_3 and PAHN_15 plates.

Neutron time interval distributions for all the Pu samples are presented in Figure 6. It can be noticed that a significant difference in the time response for the various number of Pu plates occurs at the earlier times. Neutrons emitted simultaneously in SF of $^{240}$Pu are correlated in a short time interval and therefore can be detected by organic scintillators in a coincidence time window (prompt window) of 40 ns. The later part of time distribution between 40 ns and 10 µs is due to induced fission and accidental events detected after 10 µs are due to coincidence detections of neutrons from a source and neutron background. Figure 6 shows that the slopes of the neutron time interval distributions increase when the number of PAHN Pu plates increases. The difference in the slopes of time distributions between three and five Pu plates is small. This is because the neutron multiplication factor does not change exceedingly for the Pu samples contining three and five metal plates. The intensities of time distributions depend not only on the amount of Pu in the sample but also on the source-detectors configurations.

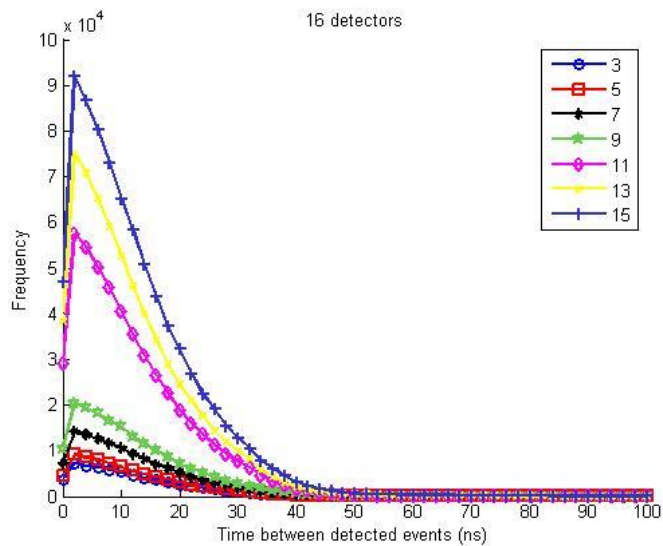

**Figure 6** Time interval distributions between neutron events for all the PAHN Pu plates with 16 detectors.

Taking into account that SF is a dominant process in a coincidence time window of 40 ns while within a wider time interval up to 10 µs IF takes place in addition to SF, we have investigated KC complexity of the time series in both time intervals. The results obtained for a short time window of 40 ns (Figure 7) and for a wider time interval (Figure 8) show that there is a clear distinction between the two data sets due to the different values of their KC.

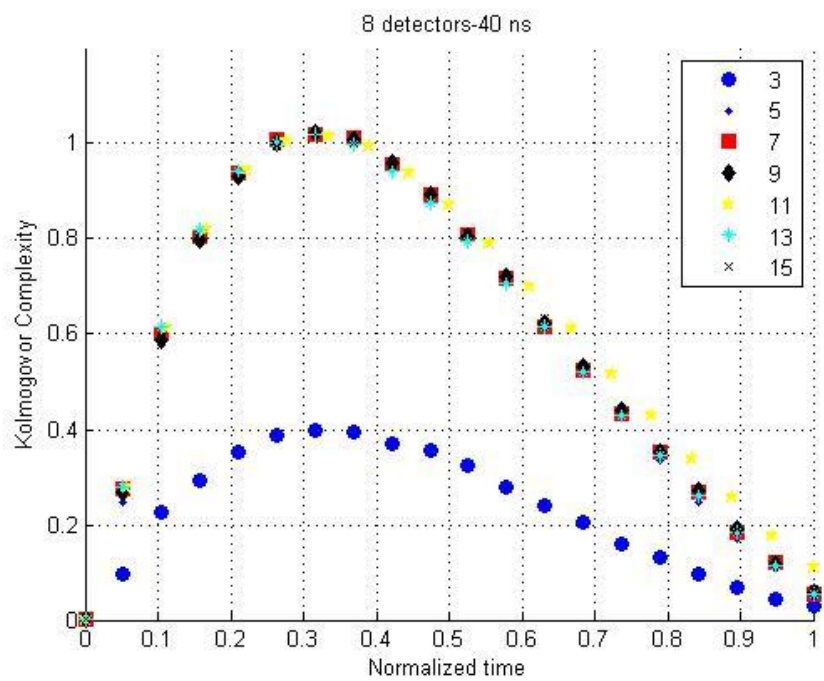

**Figure 7** Kolmogorov complexity for the time series within a short coincidence window of 40 ns.

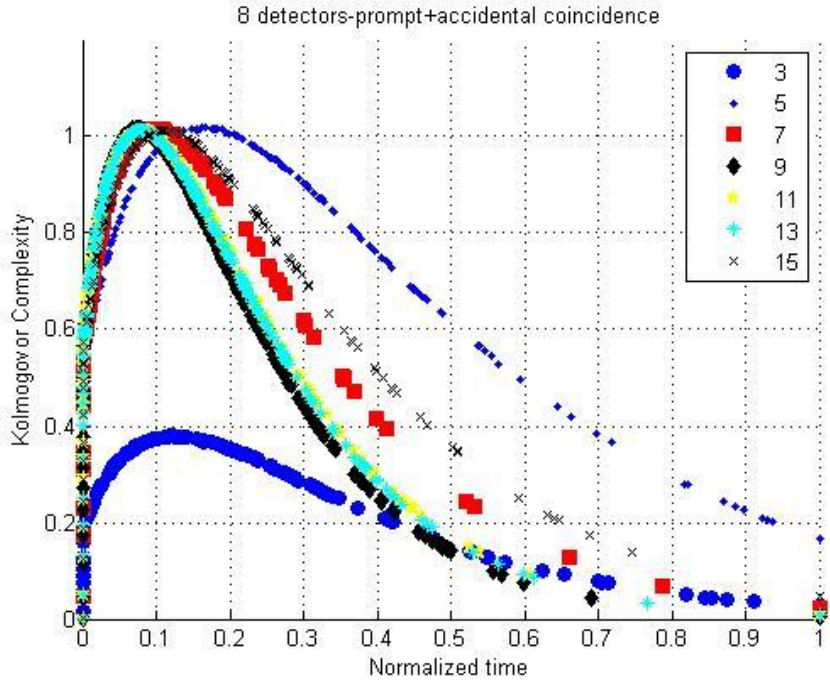

**Figure 8** Kolmogorov complexity for the time series within a wider time interval up to 10 µs.

**4.2 The randomness of time series quantified by KC complexity and its derivatives**

Complexity measures are used in a number of applications including extraction of information from data such as environmental time series (cosmic rays, solar and UV radiation, river streamflow), biomedical signals (brain and heart activity, epilepsy, schizophrenia), testing of random number generators, etc. Measuring the complexity of a time series can give crucial insights into the functioning of the system under investigation (here, a focus will be set on KC complexity and its derivatives calculated by the LZA algorithm). Let us count some examples: (i) The high value of KC in measured solar radiation time series points out the presence of very changeable cloudiness (Mihailović et al., 2018). (ii) Low values of KC for streamflow time series of turbulent rivers are often a result of human activity (Mihailović et al., 2014). (iii) In the biomedical signal analysis, a high value of KC of measurements (heart-rate, respiration rate, and blood oxygen concentration) might indicate a particular stage of sleep-apnea (Nagaraj et al., 2013). (iv) If the KC is higher the neuron activity will be more disordered; thus, the decreased KC complexity in the tasking state might be related to the increase in neuronal activity synchronization (Thilakvathi et al., 2017). However, different complexity measures proposed in the literature are ineffective in analyzing short time series that are further corrupted with noise. Note, that in our case the KC complexity is the right choice since we analyze the level of randomness for time series of very large lengths. We have computed Kolmogorov complexity (KC) and its derivatives (Kolmogorov spectrum and KC spectrum highest value - KCM) for 8 and 16 detectors, which are given in Tables 3 and 4 and shown in Figs. 9a and 9b. Inspection of Tables 3 and 4 could be summarized as follows.

*Neutrons*. (1) KC complexities are very high for both clusters. For 5, 7, 9, 11, 13 and 15 PAHN Pu plates the averages are (i) KC =1.003 (SD = 0.007 for 8 detectors) and KC = 1.001 (SD = 0.003 for 16 detectors) and (ii) KCM =1.022 (SD = 0.006 for 8 detectors) and KCM = 1.014 (SD = 0.002 for 16 detectors). (2) For 3 PAHN Pu plates there is a sharp drop of complexities KC and KCM: (i) 8 detectors (KC = 0.394, KCM = 0.398 with the drops of -60.7% and -61.1, respectively, related to the average values) and (ii) 16 detectors (KC = 0.377, KCM = 0.378 with the drops of -62.3% and -62.7%, respectively). (3) Position of KCM is placed at the same value of 12.0 ns (8 detectors), while for 16 detectors it lies within the interval [9.5, 9.9 ns]. It can be noticed that the results obtained are almost the same for 8 and 16 detectors with the samples including 5, 7, 9, 11, 13 and 15 PAHN Pu plates. Only difference is position of KCM for 8 detectors that is shifted to the right in comparison to KCM for 16 detectors (Figs. 9a and 9b).

*Gamma rays*. (1) KC complexities are very high for both clusters. For 5, 7, 9, 11, 13 and 15 PAHN Pu plates the averages are (i) KC = 0.957 (SD = 0.002 for 8 detectors) and KC = 0.981

(SD = 0.013 for 16 detectors) and (ii) KCM = 1.029 (SD = 0.004 for 8 detectors) and KCM = 1.020 (SD = 0.004 for 16 detectors). (2) A sharp drop of complexities KC and KCM for 3 PAHN Pu plates is pronounced: (i) 8 detectors (KC = 0.357, KCM = 0.382 with the drops of -62.7% and -62.9, respectively, related to the average values) and (ii) 16 detectors (KC = 0.346, KCM = 0.371 with the drops of -62.9% and -63.6%, respectively). (3) Position of KCM lies in the intervals [8.1, 10.1 ns] and [6.1, 8.4 ns] for 8 and 16 detectors, respectively. Similar to neutron statistics the gamma rays statistics is nearly the same for 8 and 16 detectors with the samples including 5, 7, 9, 11, 13 and 15 PAHN Pu plates. In contrast to peaks of KC complexity spectra for neutrons the KCM of the gamma-rays is shifted a little bit to the left (Figs. 9a and 9b).

From these basic statistics, it could be concluded that neutron and gamma events from the Pu metal plates measured either by 8 stilbene detectors either by 16 detectors (consisting of 8 stilbene detectors and 8 EJ-309 scintillators) show nearly the same results for the level of their randomness calculated using KC and KCM. These values are very high which could have been expected since it is a sign that the detectors reliably measured neutron and gamma events, i.e. without a drop in complexity. A similar conclusion can be drawn from the basic statistics of the gamma rays. However, for the 3 PAHN Pu plates, there is an evident drop of complexities for detected neutron and gamma events.

*Table 3 The Kolmogorov complexity (KC) and Kolmogorov complexity spectrum highest value (KCM) with its position in spectrum (see Figure 9a) for Pu time series measured with 8 stilbene detectors.*

| Number of Pu plates | 8 detectors | | | | | |
|---|---|---|---|---|---|---|
| | Neutrons | | | Gamma rays | | |
| | KC | KCM | Position of KCM (ns) | KC | KCM | Position of KCM (ns) |
| 3 | 0.394 | 0.398 | 12.0 | 0.357 | 0.382 | 8.1 |
| 5 | 1.004 | 1.028 | 12.0 | 0.958 | 1.030 | 8.1 |
| 7 | 1.011 | 1.028 | 12.0 | 0.961 | 1.031 | 8.1 |
| 9 | 1.010 | 1.025 | 12.0 | 0.957 | 1.033 | 8.1 |
| 11 | 0.994 | 1.020 | 12.0 | 0.957 | 1.031 | 8.1 |
| 13 | 0.997 | 1.020 | 12.0 | 0.955 | 1.030 | 8.1 |
| 15 | 1.001 | 1.014 | 12.0 | 0.956 | 1.022 | 10.1 |

*Table 4* The Kolmogorov complexity (KC) and Kolmogorov complexity spectrum highest value (KCM) and its position in spectrum (see Fig. 9b) for neutron and gamma time series measured with 16 scintillators (8 EJ-309 scintillators and 8 stilbene scintillators).

| Number of plates | 16 detectors | | | | | |
|---|---|---|---|---|---|---|
| | Neutrons | | | Gamma rays | | |
| | KC | KCM | Position of KCM (ns) | KC | KCM | Position of KCM (ns) |
| 3 | 0.377 | 0.378 | 9.9 | 0.346 | 0.371 | 6.1 |
| 5 | 1.005 | 1.017 | 9.5 | 0.961 | 1.015 | 6.1 |
| 7 | 1.000 | 1.015 | 9.9 | 0.981 | 1.018 | 8.4 |
| 9 | 1.005 | 1.014 | 9.5 | 0.987 | 1.021 | 8.4 |
| 11 | 0.998 | 1.012 | 9.5 | 0.991 | 1.026 | 8.4 |
| 13 | 1.000 | 1.012 | 9.5 | 0.996 | 1.021 | 8.4 |
| 15 | 1.000 | 1.012 | 9.5 | 0.972 | 1.018 | 8.4 |

Although the KC complexity is a general indicator of randomness this drop indicates a process or event which interferes with the functioning of detectors or limitation of the measuring system.

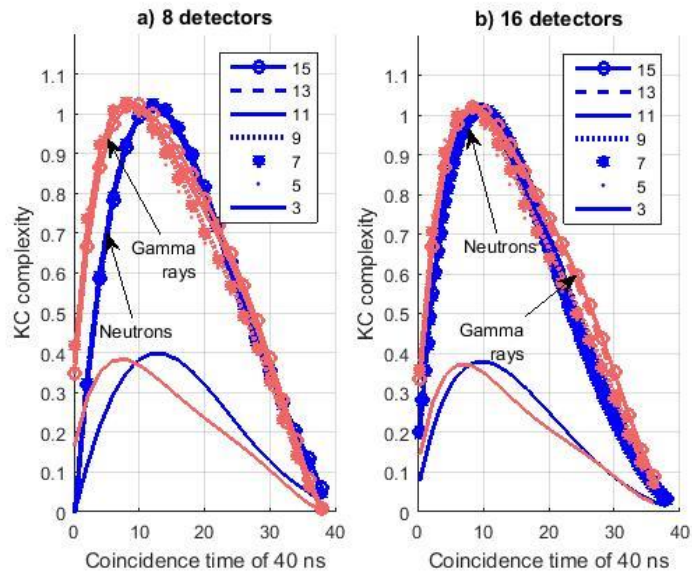

**Figure 9** KC spectra for neutrons and gamma rays for the Pu time seriesmeasured with 8 and 16 detectors.

**4.3 The sharp drop of complexity ("breaking point") in highly random Pu time series**

The problem of defining and studying the complexity of a time series has the scientific community interested for years. Mostly, it was in the context of getting predictive information through information measures. Usually, KC complexity and Shannon has been used. Thus, Bialek et al. (2001) pointed out that our intuitive notion of complexity corresponds to statements about the underlying process, and not directly to Kolmogorov complexity. A dynamic process with an unpredictable and random output (large KC) may be as small as the dynamics producing the predictable constant outputs (small KC) – although "really" complex processes lie somewhere in between (Prokopenko et al., 2009). This fact is visualized by Crutchfield and Young (1990) via the complexity spectrum in Figure1 where is drown complexity $C$ as a function of the diversity of patterns against the normalized Shannon entropy (Fig. 10). Regular data have low entropy; very random data have maximal entropy. However, their complexities are both low.

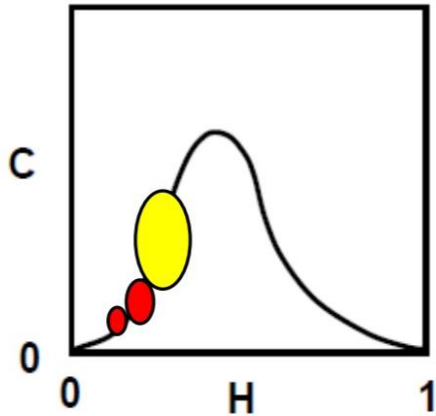

**Figure 10** The complexity spectrum: complexity *C* as a function of the diversity of patterns. The latter is measured with the (normalized) Shannon entropy *H* (Crutchfield and Young, 1990). The different areas of ellipses symbolically show a transition of the Pu time series complexity level from high complexity (yellow) and low Shannon entropy (and also low KC) towards the "breaking point" and approaching to zero (red ellipses).

The comments by Crutchfield and Young (1990) and Figure 10 could be a good slogan for commenting on the occurrence of the "breaking point" of running complexity of the neutron and gamma PAHN_3 time series. All the Pu time series have thelowervalues of normalized Shannon entropy while the KC values are very high (Tables 3 and 4). From Figure 11 is seen that the running complexity of PAHN_7 is nearly constant (it is the same for time series of higher number with Pu plates, which are not drawn), while for PAHN_3 time series a "breaking point" is occurred. This situation we can attribute to the neutron and gamma-rays interaction with detector rather than to the nuclear process itself. Namely, the architecture of detectors configuration and influence of surrounding environment on the detector can introduce some uncertainties in the measuring procedure. Thus, all the Pu time series (except PAHN_3) are measured reliably. Complexities of these time series lie in the (C, H) space symbolically covered by the yellow ellipse (Figure 11). In contrast to that, the PAHN_3 time series have high KC complexities and then reaches a point ("breaking point") after which they quickly go down.

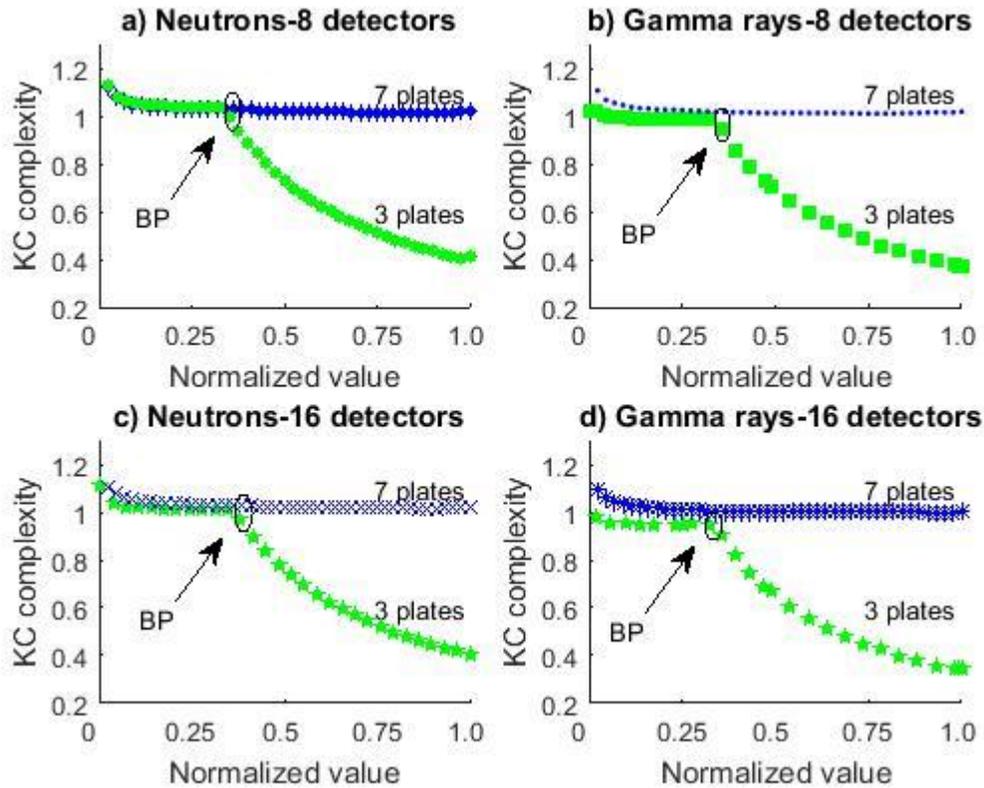

**Figure 11** The running complexity of the neutron and gamma time series with 3 and 7 PAHN plates. The "breaking points" are rounded by ellipses. The position of sample on the x – axis is normalized by its the length of a time series. BP is notation for "breaking point".

We have shown that for 3 PAHN plates there is a drop of KC, same as in the logistic map, with $r= 3.95$ (Figs. 11 and Fig. 2). From Fig. 2 is seen that Lyapunov exponent is positive for all $r$ values indicating the chaotic nature of the time series. The results of calculating the Lyapunov exponent of Pu time series are shown in Fig.12. For both, gamma rays and neutrons, values of the

Lyapunov exponent of PAHN_5-PAHN_15 time series are placed in the interval (0.45, 0.55) characterizing the high chaos.

All Pu time series have high KC complexity with an exception for PAHN_3 plate, whose time series (gamma and neutron) have Lyapunov exponent in the region of the high chaos and lower value of the KC (isolated island in the left side of Fig. 12). This is not contradictory since the complexity is collective dynamical behavior from simple interactions between large numbers of subunits while chaos is the generation of complicated, aperiodic, apparently random behavior from the iteration of a simple rule. "Stand alone chaos and complexity have absolutely nothing to do with generating formal function. Chaotic systems are not necessarily complex, and complex systems are not necessarily chaotic" (Rickles et al., 2007; Bertuglia, and Vaio F, 2005).

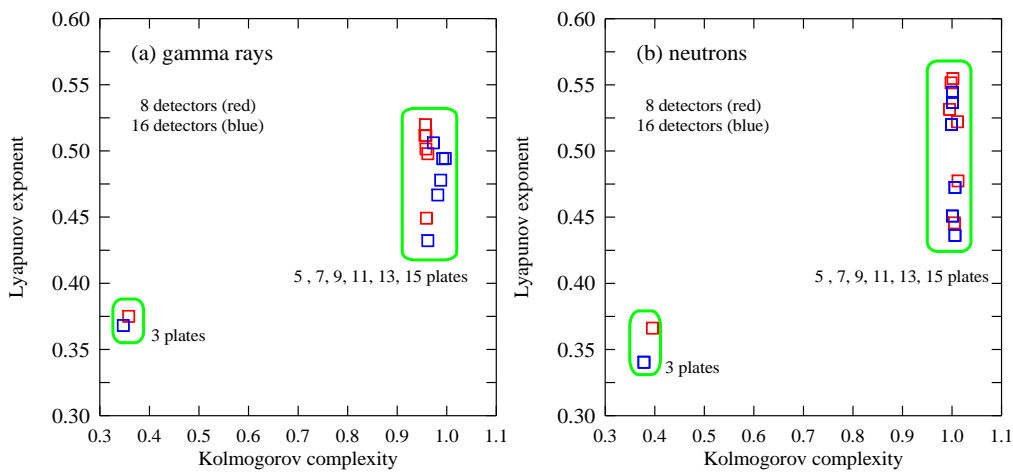

**Figure 12** The Lyapunov exponent of the neutron and gamma time series for all Pu plates as a function of KC complexity.

We have performed NIST tests for 28 time series [7 (gamma rays) and 7 (neutrons) for 8 and 16 detectors]. The test having $p > 0.01$ has been noted as passed; otherwise notation was no-passed. All time series passed this test except for PAHN_3 and PAHN_5 time series (for both 8 and 16 detectors). It means that these time series are no-random. Having in mind these facts, we can say that the PAHN_3 and PAHN_5 time series are chaotic because of the nature of neutron and gamma radiation. Lower KC complexity and no-random nature (relative to other time series) of PAHN_3 time series can be attributed to the different sources of uncertainties in detection procedure with the sample consisting of three plates.

## 5. Conclusions

One of the measures to investigate the reliability of the Pu time series measurement was used to obtain a general estimate of the complexity of Pu time series. To access this information, we applied Kolmogorov complexity (KC) and running KC complexity to 28-time series (14 measured with 8 detectors and 14 measured with 16 detectors) each including an odd number of Pu plates from 3 to 15.

We obtained several novel and interesting findings. (1) By means of numerical experiments with LE (logistic parameter took values in the interval [3.145, 3.155]) we established the existence of so called "breaking point" in RKC, i.e., a point at which the KC complexity of the LE time series have a very sharp drop in complexity. (2) The running complexity is capable of revealing aspects of information that would otherwise remain hidden to the one-off complexity estimates. Even when difficult to apply, it can be useful in disentangling complex phenomena of detection of neutrons and gamma-rays emitted in a short time interval. As confirmed in previous studies, KC produces excellent results with running complexity as long as the window is sufficiently large to accommodate the measure's requirements. (3) We discovered a presence of the "breaking point" in the running KC complexity of the Pu three plates time series (PAHN_3). After the "breaking point" they have a very pronounced drop in running complexity (i) for neutron and gamma rays and (ii) for both clusters with 8 and 16 detectors. (4) From this statistics, it could be concluded that neutron and gamma-rays from Pu samples measured either by 8 or 16 detectors show nearly the same results for their very high randomness, which is calculated using KC complexity and the highest value of in the KC spectra (KCM). It is a sign that the detectors reliably measured the neutron and gamma events on the nanosecond time-scale, i.e. without any drop in their complexity. (5) The running complexity of the PAHN_3 time series has the "breaking point". This situation can attribute to the sensitivity and stability of the measurement equipment during the measurement procedure as well as to the contribution of radiation background which was not constant during the acquisition time rather than to the nuclear process itself.

**Acknowledgment**

The authors are grateful to the authorities of the Detection for Nuclear Nonproliferation Group (DNNG) at the University of Michigan for giving us an opportunity to have access to the neutron and gamma time series. The authors are especially grateful to the anonymous reviewer who with implicit and explicit comments contributed to the credibility of the paper. Prof. M. Sys from the Masaryk University in Brno gave us precise and valuable comments. Ms Emilija Nikolić – Đorić was an excellent advisor in the statistics.

**References**

Bertuglia, C.S., Vaio F., 2005: Nonlinearity, chaos and complexity: The dynamics of natural and social systems. Oxford University Press, Oxford (UK).

Bialek, W., Nemenmana, I., Tishbya, T., 2001: Complexity through nonextensivity, Physica A 302, 89–99.

Chichester, D.L., S.J. Thompson, M.T. Kinlaw, J.T. Johnson, J.L. Dolan, M. Flaska, S.A. Pozzi, 2015:Statistical estimation of the performance of a fast-neutron multiplicity system for nuclear material accountancy, Nucl. Instrum. Methods Phys. Res. A. 784, 448–454.

Clarke, S.D., Flaska, M., Pozzi, S.A., Peerani, P., 2009: Neutron and gamma-ray cross-correlation measurements of plutonium oxide powder, Nuclear Instruments and Methods in Physics Research Section A Accelerators Spectrometers Detectors and Associated Equipment 604(3):618-623, DOI: 10.1016/j.nima.2009.02.045.

Crutchfield, J.P. and Young, K., 1990: Computation at the onset of chaos. In W. H. Zurek, editor, Complexity, Entropy, and the Physics of Information, pages 223-269. Addison Wesley, Redwood City, CA.

Di Fulvio, A., Shin, T.H., Jordan, T., Sosa, C., Ruch, M.L., Clarke, S.D., Chichester, D.L., Pozzi, S.A., 2017: Passive assay of plutonium metal plates using a fast-neutron multiplicity counter Nuclear Instruments and Methods in Physics Research A 855, 92–101.

Duhem, P., 1954: The aim and structure of physical theory (Foreword by Prince Louis de Broglie). Princeton University Press, Princeton, N.J.

Ensslin, N., Harker, W.C., Krick, M.S., Langner, D.G., Pickrell, M.M., Stewart, J.E., 1998: Application Guide to Neutron Multiplicity Counting, LA-13422-M, Los Alamos National Laboratory.

http://www.matjazperc.com/time/

Khrennikov; A., 2014: Introduction to foundations of probability and randomness (for students in physics), Lectures given at the Institute of Quantum Optics and Quantum Information, Austrian Academy of Science, Lecture-1: Kolmogorov and von Mises. arXiv-qunt-ph.1410.5773

Knoll, G.F., 2010: Radiation Detection and Measurement, 4th ed, John Wiley & Sons, Hoboken.

Kodba, S., Perc, m. and Marhl, m., 2005: Detecting chaos from a time series, Eur. J. Phys. 26, 205-215.

Kolmogorov, A., 1965: Three approaches to the quantitative definition of information." Problems of Information Transmission, 1 p. 4.

Kovalsky, M., Hnilo, A., (2018) Agüero, M.B. Kolmogorov complexity of sequences of random numbers generated in Bell's experiments. ArXiv, 2018; arXiv:1805.07161.

Lempel, A., Ziv, J. (1976). On the complexity of finite sequences. *IEEE Transactions on Information Theory, 22*, 75–81.

Mihailović, D., Mimić, G., Gualtieri, P., Arsenić, I., Gualtieri, C., 2017: Randomness representation of turbulence in canopy flows using Kolmogorov complexity measures. *Entropy*, *19*, 519.

Mihailović, D.T., Bessafi, M., Marković, S., Arsenić, I., Malinović-Milićević, S., Jeanty, P., Delsaut, M., Chabriat, J.-P., Drešković, N., Mihailović, A., 2018: Analysis of solar irradiation time series complexity and predictability by combining Kolmogorov measures and Hamming Distance for La Reunion (France). Entropy, 20, 57

Mihailović, D.T., Mimić, G., Nikolić-Djorić, E., Arsenić, I., 2015: Novel measures based on the Kolmogorov complexity for use in complex system behavior studies and time series analysis. Open Phys 13:1-14

Mihailović, D.T., Nikolić-Đorić, E. Drešković, N. and Mimić, G., 2014, Complexity analysis of the turbulent environmental fluid flow time series, Physica A 395, 96-104.

Mihailović, D.T., Nikolić-Đorić, E., Arsenić, I., Malinović-Milićević, S., Singh, V. P., Stošić, T., 2019: Analysis of daily streamflow complexity by Kolmogorov measures and Lyapunov exponent, Physica A 525, 290-303.

Mihailović, D.T., Nikolić-Đorić, E., Malinović-Milićević, S., Singh, V.P., Mihailović, A., Stošić, T., Stošić, B., Drešković, N., 2019: The Choice of an Appropriate Information Dissimilarity Measure for Hierarchical Clustering of River Streamflow Time Series, Based on Calculated Lyapunov Exponent and Kolmogorov Measures. *Entropy, 21*, 215.

Mimić, G., Arsenić, I., and Mihailović, D. T., 2016: Number of Patterns in Lempel-Ziv Algorithm Applied to Iterative Maps and Measured Time Series, International Congress on Environmental Modelling and Software. 63. https://scholarsarchive.byu.edu/iemssconference/2016/Stream-A/63

Mimić, G., Arsenić, I., Mihailović, D., 2016: Number of Patterns in Lempel-Ziv Algorithm Applied to Iterative Maps and Measured Time Series. In: Sauvage, S., Sánchez-Pérez, J.M., Rizzoli, A.E. (Eds.), Proceedings of the 8th International Congress on Environmental Modelling and Software, July 10-14, Toulouse, FRANCE. ISBN: 978-88-9035-745-9

Nagaraj, N., Balasubramanian, K. & Dey, S., 2013: A new complexity measure for time series analysis and classification. Eur. Phys. J. Spec. Top. 222**,** 847–860.

Nakae, L., Chapline, G., Glenn, A., Kerr, P., Kim, K., Ouedraogo, S., Prasad, M., Sheets, S., Snyderman, N., Verbeke, J.,Wurtz**,** R., 2011:Recent Developments In Fast Neutron Detection And Multiplicity Counting With Verification With Liquid Scintillator, October 4, 2011**,** LLNL-CONF-502912.

Oberstedt, S., Oberstedt, A., Gatera, A., Göök, A., F.-J. Hambsch, F.-J., Moens, A., Sibbens, G., Vanleeuw, D. and Vidali, M., 2016: Prompt fission γ-ray spectrum characteristics from 240Pu(sf) and 242Pu(sf), Phys. Rev. C 93, 054603

Pazsit, I., Pozzi, S.A., 2015: Calculation of gamma multiplicities in a multiplying sample for the assay of nuclear materials, Nuclear Instruments and Methods in Physics Research A 555 340–346

Perez, C.E, 2017: https://intuitmachine.medium.com/

Polack, J.K, Flaska, M., Enqvist, A., Sosa, C.S., Lawrence, C.C., Pozzi, S.A., 2015: Analgorithm for charge-integration, pulse-shape discrimination and estimation ofneutron/photon misclassification in organic scintillators, Nucl. Instrum. Methods Phys. Res. A. 795, 253–267.

Prokopenko, M., Boschetti, F., Ryan, A.J., 2009: An information-theoretic primer on complexity, self-organisation and emergence, Complexity 15 (1), 11–28.

Radhakrishnan, N., Wilson, J.D., Loizou, P.C., 2000: An alternate partitioning technique to quantify the regularity of complex time series. Int J Bifurcat Chaos 10:1773–1779

Rickles, D., Hawe, P., &Shiell, A. (2007). A simple guide to chaos and complexity. Journal of epidemiology and community health, 61(11), 933–937. https://doi.org/10.1136/jech.2006.054254

Robinson, S.M, Runkle,R.C., Newby, R.J., 2011: A comparison of performance between organic scintillation crystals and moderated 3He-based detectors for fission neutron detection, Nucl. Instrum. Methods Phys. Res. A. 652, 404–407.

Shen, V.K., Siderius, D.W., Krekelberg, W.P., and Hatch, H.W. (2016) Eds., NIST Standard Reference Simulation Website, NIST Standard Reference Database Number 173, National Institute of Standards and Technology, Gaithersburg MD, 20899, http://doi.org/10.18434/T4M88Q

S.P.A. CAEN, V1730 user manual, 2016.

Sys M. , Riha Z., Maty V., Marton K., Suciu A. (2015) On the interpretation of results from the NIST statistical test suite. Romanian J. Inf. Sci. Technol. 18, 18-32.

Thilakvathi B., Bhanu K., Malaippan M., 2017 : EEG signal complexity analysis for schizophrenia during rest and mental activity. Biomed. Res. 28, 1–9.

Vaughan, S., 2013: Random time series in astronomy. Philos. Trans. R. Soc. A. 371, 20110549

Zucker, M.S., Holden, N.E., 1986: Energy Dependence of Neutron Multiplicity P(ν) in Fast-Neutron-Induced Fission for 235,238U and 239Pu, BNL-38491.